\documentclass[12pt,preprint]{aastex}
\usepackage{natbib}
\usepackage{psfig}
\usepackage{epsfig}
\usepackage{graphicx}
\title{A Neutral Hydrogen Self-Absorption Cloud in the SGPS}
\author{D. W. Kavars}
\affil{Department of Astronomy, University of Minnesota, 116 Church St. SE, Minneapolis, MN, 55455, dkavars@astro.umn.edu}
\author{J. M. Dickey}
\affil{Department of Astronomy, University of Minnesota, 116 Church St. SE, Minneapolis, MN, 55455, john@astro.umn.edu}
\author{N. M. McClure-Griffiths}
\affil{Australia Telescope National Facility, CSIRO, PO Box 76, Epping NSW 1710, Australia, Naomi.McClure-Griffiths@atnf.csiro.au}
\author{B. M. Gaensler}
\affil{Harvard-Smithsonian Center for Astrophysics, 60 Garden St. MS-6, Cambridge, MA, 02138, bgaensler@cfa.harvard.edu}
\author{A. J. Green}
\affil{Astrophysics Department, School of Physics, University of Sydney, Sydney, NSW 2006, Australia, agreen@physics.usyd.edu.au}

\begin{abstract}
Using data from the Southern Galactic Plane Survey (SGPS) we analyze an HI self-absorption cloud centered on l = 318.0$^{\circ}$, b = -0.5$^{\circ}$, and velocity, v = -1.1 km s$^{-1}$.  The cloud was observed with the Australia Telescope Compact Array (ATCA) and the Parkes Radio Telescope, and is at a near kinematic distance of $\lesssim$ 400 pc with derived dimensions of $\lesssim$ 5 $\times$ 11 pc.  We apply two different methods to find the optical depth and spin temperature.  In both methods we find upper limit spin temperatures ranging from 20 K to 25 K and lower limit optical depths $\sim$ 1.  We look into the nature of the HI emission and find that 60-70\% originates behind the cloud.  We analyze a second cloud at the same velocity centered on l = 319$^{\circ}$ and b = 0.4$^{\circ}$ with an upper limit spin temperature of 20 K and a lower limit optical depth of 1.6.  The similarities in spin temperature, optical depth, velocity, and spatial location are evidence the clouds are associated, possibly as one large cloud consisting of smaller clumps of gas.  We compare HI emission data with $^{12}$CO emission and find a physical association of the HI self-absorption cloud with molecular gas.  
\end{abstract}

\begin{document}


\section{Introduction}

Cold clouds of neutral hydrogen (HI) in the interstellar medium (ISM) are difficult to observe and our knowledge of them is limited.  In contrast to the warmer but still cool HI clouds with spin temperatures of 60-80 K \citep{1971ApJS...23..323H, 1972ApJS...24...15R, 1988gera.book...95K}, cold HI clouds are found with spin temperatures as low as 10 K \citep{2001Natur.412..308K}.  Other studies find cold HI with spin temperatures of 8-36 K \citep{1980ApJ...242..416L}, 35 K \citep{1995MNRAS.273..449M}, and 15-60 K \citep{2000ApJ...540..851G}.  Cold HI is detected by looking at 21cm emission profiles. If there is a cold cloud in front of warmer background emission, an absorption dip is present in the emission profile.  This is known as HI self-absorption (HISA).  

HI self-absorption was known as early as 1955 \citep{1955ApJ...121..569H}.  Unfortunately variations in HI emission cause emission profiles to be quite complex, leaving the self-absorption dips difficult to distinguish from valleys.  \citet{1974AJ.....79..527K} observed 88 dark dust clouds, finding spin temperatures of 15-40K, and developed the following criteria for self-absorption:  (1) The absorption is narrow with a width no larger than a few km s$^{-1}$.  (2) The HI velocity is in agreement with velocities of any molecular lines observed. (3) There should be no corresponding dip in the off-cloud profile and (4) the steepness of the dip should be greater than the background emission profile.  This presents a few problems.  Because the absorption dips are only a few km s$^{-1}$ wide, high velocity resolution is needed to detect them.  This is achieved with single dish telescopes, but the angular resolution is lower.  To obtain higher angular resolution, interferometers are used \citep{1988A&A...201..311V, 2000ApJ...540..851G}.  The work we present in this paper uses data from the Southern Galactic Plane Survey(SGPS) \citep{2001PhDT.........9M}.  

Difficulties in detecting HI self-absorption include determining the off-cloud profile, the necessity of a bright, uniform HI background, and the capability of only setting limits on the spin temperature and optical depth of the cloud.  Determining the off-cloud profile is done by assuming a uniform background temperature \citep{2001Natur.412..308K, Li} and measuring the brightness temperature in directions just off the cloud.  If the background temperature varies significantly over the spatial extent of the cloud, there will be large uncertainties in the spin temperature.  Most studies, including this one, invoke fitting function(s) to the HI emission profile to interpolate across the absorption dip \citep{1983AJ.....88..658H, 1993A&A...276..531F, 2000ApJ...540..851G}.  If the HI emission is not bright and relatively uniform, large scale cold HI structures are difficult to detect.  This is why very few large HISA features are known. The radiative transfer equation is problematic because the spin temperature and optical depth cannot be found simultaneously.  In most cases attempts are made at constraining both and assumptions are made about one or the other. For example, \citet{2000ApJ...540..851G} assume the optical depth is no larger than 10.  The work we present here calculates an upper limit spin temperature of $\sim$ 20 K and uses this value to compute the optical depth, as one example.   

Although HISA clouds have proven difficult to detect, their relationship with other components of the ISM is very important. \citet{1978ApJ...219L..67B} show that molecular clouds could contain cold atomic hydrogen which results in narrow HI self-absorption features.  Molecular clouds consist primarily of molecular hydrogen, but there are still  traces of atomic hydrogen due to dissociative processes \citep{Li}.  With a mix of H$_2$ and HI in the cloud, H$_2$ can be observed by using $^{12}$CO or $^{13}$CO as a tracer. Thus, there should be a spatial correlation between $^{12}$CO or $^{13}$CO emission and HI self-absorption.  \citet{1981ApJ...246...74L} provide observational support for this correlation and demonstrate that molecular clouds can dramatically effect HI line profiles at low latitudes.  These effects include small-scale scatter in measured HI terminal velocities, general lumpiness in the emission structure of HI and a blanketing effect on HI intensities observed near the terminal velocity.  \citet{1984A&A...140..303L} run a three dimensional simulation of a molecular cloud containing cold HI and find the largest effect is that the opacity of the material blankets and distorts the high-velocity portions of inner Galaxy HI profiles.  \citet{1979A&AS...35..129B} search for $^{12}$CO emission in six HI self-absorption clouds and find a strong spatial correlation in each.  Other studies showing this include those by \citet{1981ApJ...243..778L} and \citet{1988A&A...207..145J}.  However, the  more recent study by \citet{2000ApJ...540..851G} reveals a discrepany between $^{12}$CO and HISA, with only a fraction of HISA clouds containing detected $^{12}$CO. This questions our current knowledge of molecular clouds and demonstrates the need for continued high resolution studies of HISA and $^{12}$CO. 

This paper will show that this HISA cloud has a spin temperature and optical depth similar to that of the past studies listed above.  However, it is the appearance of the cloud on the sky that sets it apart from other clouds.  Although its physical size is not particularly large, it spans 1$^\circ$ $\times$ 1.5$^\circ$ on the sky.  With a high resolution survey such as the SGPS, we can easily observe the structure of the cloud both spatially and in velocity, whereas past surveys have had difficulties doing this.  This becomes crucial when comparing the cloud to $^{12}$CO emission and determining the physical processess involved.  

In this paper we present an analysis of a prominent self-absorbing neutral hydrogen cloud in the Southern Galactic Plane Survey.  In Section 2 we briefly discuss the SGPS data.  In Section 3 we use a four component radiative transfer model and present two methods of analyzing the HISA cloud.  In Section 4 we analyze a second, nearby HISA cloud, and in Section 5 we compare the cloud to $^{12}$CO emission and discuss our results.

\section{Data}

The Southern Galactic Plane Survey images HI line emission and 21cm continuum in the fourth quadrant of the plane of the Galaxy \citep{2001PhDT.........9M}.  The SGPS combines data from the Australia Telescope Compact Array (ATCA) and the Parkes 64m single dish.  It maps the region of $253^{\circ} \leq l \leq 358^{\circ}$ and $-1^{\circ} \leq b \leq  +1^{\circ}$ with angular resolution of 2$^{\prime}$ and velocity resolution of 0.82 km s$^{-1}$, with single dish data extending to b = $\pm$10$^{\circ}$.  The SGPS is valuable in tracing HI shells and large HI supershells \citep{2000AJ....119.2828M, 2001ApJ...562..424M}, but it can also detect HISA features \citep{2001PASA...18...84M}.  With data from an interferometer and a single dish telescope, the SGPS provides higher angular resolution than a single dish alone.  Although the velocity resolution of the SGPS is lower than some single dish surveys \citep{1974AJ.....79..527K, 1978MNRAS.185..755M}, it is still high enough to detect the narrow features seen in self-absorption profiles.  The HI self-absorption cloud we present in this paper at $l = 318.0^{\circ}$, $b = -0.5^{\circ}$, and $v = -1.1$ $km$ $s^{-1}$, is shown in Figure 1 as eight velocity channel images.  It is most prominent in the v = -0.82 km s$^{-1}$ channel.  The cloud spans $\sim$ 1$^\circ$ $\times$ 1.5$^\circ$ on the sky and has a velocity line width of $\sim$ 3 km s$^{-1}$.  There is a second cloud possibly associated with this cloud at l = 319$^{\circ}$, b = 0.4$^{\circ}$, and v = -1.1 km s$^{-1}$.  This is discussed further in Section 4.  We estimate the distance to the first cloud by, 

\begin{equation}
d = \frac{V_R}{A*\sin{(2l)}}
\end{equation}

\noindent where $V_R$ is the radial velocity of the cloud, $A$ is Oort's constant($14.5$ $km$ $s^{-1}$ $kpc^{-1}$), and $l$ is the longitude of the cloud.  The cloud to cloud velocity dispersion is approximately 6.9 $km$ $s^{-1}$ \citep{1984A&A...136..368B}.  Assuming the velocity profile is a gaussian centered about -1.1 $km$ $s^{-1}$ and setting one standard deviation equal to 6.9 $km$ $s^{-1}$, a probability can be set on the distance.  Equation 1 is good out to a few kiloparsecs \citep{1998ggs..book.....E} and gives an upper limit distance of $\lesssim$ 400 pc at the 68\% confidence level.  At the 95\% confidence level, the upper limit distance is $\lesssim$ 900 pc.  Because HISA is seen, the cloud is probably at the near-kinematic distance \citep{2002ApJ...566L..81J}.  At a distance of 400 pc, the upper limit spatial dimensions are $\lesssim$ 5 $\times$ 11 pc.  A velocity profile is shown in Figure 2.  The brightness temperature abruptly drops and quickly increases following the center of the dip at -1.1 $km$ $s^{-1}$.  The steepness of the dip and the narrow line width are evidence that this is a self-absorption cloud.  A second velocity profile with a broader range in velocity is shown in Figure 3.  The HISA cloud is represented by the sharp dip in the profile near v = 0 km s$^{-1}$.  The continuum map in Figure 4 shows a number of compact sources.  Catalogued sources \citep{1987A&A...171..261C, 1996A&AS..118..329W} are marked.  HII regions and SNR's are labeled by +'s and X's, respectively.

\section{Analysis}

\subsection{Radiative Transfer}

We find the optical depth and spin temperature of the HISA cloud using the radiative transfer equation, 

\begin{equation}
I_{\nu} =  I_{\nu,max}(\tau_{max})e^{-\tau_{max}} + \int_0^{\tau_{max}} B_{\nu}(\tau_{\nu})e^{-\tau_{\nu}}d\tau_{\nu}
\end{equation}

\noindent where $I_{\nu}$ is the observed intensity, $I_{\nu{,max}}$ is the intensity incident behind the HISA cloud, and $B_{\nu}$ is the source function, $j_{\nu}/\kappa_{\nu}$.  The emission coefficient is $j_{\nu}$ and $\kappa_{\nu}$ is the absorption coefficient.  The optical depth along the line of sight is denoted by $\tau_{\nu}$ with a maximum $\tau_{max}$.  Changing from intensities to brightness temperatures and using a four component model similar to that of \citet{2000ApJ...540..851G} consisting of a HISA cloud, foreground and background HI emission, and four sources of diffuse continuum shown in Figure 5, Equation 2 becomes:

\begin{eqnarray}
T_{on} &=& T_{c,A} + [T_s]_{fg}(1 - e^{-{\tau}_{fg}}) + T_{c,B}e^{-{\tau}_{fg}} \nonumber\\
	&+& [T_s]_{HISA}(1 - e^{-{\tau}_{HISA}})e^{-{\tau}_{fg}} + T_{c,C}e^{-({\tau}_{fg}+{\tau}_{HISA})} \nonumber\\
	&+& [T_s]_{bg}(1 - e^{-{\tau}_{bg}})e^{-({\tau}_{fg} + {\tau}_{HISA})} + T_{c,D}e^{-({\tau}_{fg} + {\tau}_{HISA} + {\tau}_{bg})}
\end{eqnarray}

The measured on-cloud brightness temperature is $T_{on}$, the spin temperature is $T_s$, and $T_{c}$ is the measured continuum brightness temperature.  The subscripts `fg' and `bg' represent foreground and background gas, relative to the HISA cloud.   The `A', `B', `C', and `D' subscripts refer to different pieces of continuum as seen in Figure 5.  At a distance of $\lesssim$ 400 pc, there will be some foreground HI emission, but compared to the background HI emission, this will be small.  Setting ${\tau}_{fg}$ small compared to unity, assuming a small HISA cloud, and setting ${\tau}_{HISA} = 0$ to find the off-cloud brightness temperature, the difference between the on-cloud and off-cloud brightness temperatures is: 

\begin{equation}
T_{on} - T_{off} = ([T_s]_{HISA} - [T_s]_{bg}(1-e^{-{\tau}_{bg}}) - T_{c,C} - T_{c,D}e^{-{\tau}_{bg}})(1 - e^{-{\tau}_{HISA}})
\end{equation}

$T_{on}$ and $T_{off}$ are measured brightness temperatures at positions on and off the HISA cloud, respectively, as shown in Figure 5, with the continuum baseline subtracted.  Equation 4 can be simplified further by letting $T_{cont} = T_{c,C} + T_{c,D}e^{-{\tau}_{bg}}$.  $T_{cont}$ is the continuum emission from behind the HISA cloud.  We also define a new a variable, `p', where `p' describes the amount of HI emission originating behind the cloud.

\begin{equation}
p = \frac{[T_s]_{bg}(1-e^{-{\tau}_{bg}})}{[T_s]_{fg}(1-e^{-{\tau}_{fg}}) + [T_s]_{bg}(1-e^{-{\tau}_{bg}}) + T_{c,D}(e^{-{\tau}_{bg}}-1)}
\end{equation}

Substituting for $p$ and $T_{cont}$ and solving for the optical depth of the cloud gives:

\begin{equation}
{\tau}_{HISA} = -ln(1 - \frac{T_{on} - T_{off}}{[T_s]_{HISA} - T_{cont} -pT_{off}})
\end{equation}

Equation 6 gives $\tau$ as a function of $T_s$ (or vice versa), but the two cannot be derived simultaneously.  Here we are not assuming that $\tau_{bg}$ is small, but only that it does not have a narrow feature at the velocity of the HISA cloud.  Thus, we are assuming that $\tau_{bg}$ is smooth over the velocities -9 to +5 km s$^{-1}$.  An example curve is shown in Figure 6.  Barring any constraints on optical depth or spin temperature, a solution lies anywhere on the $\tau$ vs. $T_s$ curve, down to a limiting spin temperature no lower than the microwave background temperature of 2.7 K and up to a maximum spin temperature found by looking at the asymptotic value of $\tau$.  A realistic solution lies somewhere between these values, but using this analysis we only find an upper limit to the spin temperature and a lower limit to the optical depth of the HISA cloud.

\subsubsection{Method One:  Latitude Slices}

We apply two methods in our analysis of the HISA cloud.  In the first method, we analyze seventeen constant latitude slices through the cloud at intervals of $\Delta$b = $8^{\prime}$ from b = $-1.39^{\circ}$ to b = $+0.75^{\circ}$, shown in Figure 7, and $T_{on}$, $T_{off}$, and $T_{cont}$ are measured at each latitude.  For each latitude slice, a pixel at the deepest point of the slice is selected and $T_{on}$ and $T_{cont}$ are measured at this position.  To find $T_{off}$, we look at the left and right ends of the slice and take an average of the brightness temperature at the ends to obtain an estimate for $T_{off}$ at the center of the cloud.  This procedure is illustrated in Figure 8.

Using this method we obtain values for $T_{on}$, $T_{off}$, and $T_{cont}$ for each slice.  Putting these values into Equation 6 results in seventeen $\tau$ vs. $T_s$ curves.  These curves still depend on the value of $p$.  At a distance of $\lesssim$ 400 pc, the cloud is outside the Local Bubble.  The Local Bubble, stretching 50 to 100 pc from the Sun, is almost entirely ionized, although the inner 1-2 pc are warm neutral medium (WMN) \citep{2002ApJ...565..364S}.  Between the Local Bubble and the HISA cloud there is more foreground HI, but with this upper limit distance, it is likely that a majority of the HI emission originates behind the cloud.  A conservative estimate assumes half to three quarters of the HI emission originates behind the cloud(i.e. $p = 0.5$ to 0.75). We later show that 60-70\% of the HI emission originates behind the HISA cloud.  $T_{on}$, $T_{off}$, and $T_{cont}$ are found for all seventeen latitude slices and the results are shown in Table 1.

\begin{deluxetable}{ccccc}
\tablewidth{0pc}
\tablecaption{Observed parameters for seventeen latitude slices.}
\tablehead{
\colhead{Slice\#}  & \colhead{Latitude(Deg)}  & \colhead{$T_{on}(K)$}  &
\colhead{$T_{off}(K)$} & \colhead{$T_{cont}$}}
\startdata

	1 & -1.39 & 59 & 69 & 11 \\ 
	2 & -1.26 & 57 & 72 & 12 \\ 
	3 & -1.12 & 54 & 71 & 12 \\
	4 & -0.99 & 42 & 68 & 13 \\ 
	5 & -0.86 & 43 & 69 & 14 \\
	6 & -0.72 & 42 & 69 & 13 \\
	7 & -0.59 & 37 & 67 & 14 \\ 
	8 & -0.46 & 34 & 70 & 16 \\
	9 & -0.32 & 37 & 68 & 15 \\ 
	10 & -0.19 & 31 & 61 & 16 \\ 
	11 & -0.06 & 31 & 60 & 16 \\ 
	12 & 0.08 & 23 & 51 & 18 \\
	13 & 0.21 & 24 & 47 & 16 \\ 
	14 & 0.34 & 30 & 42 & 14 \\ 
	15 & 0.48 & 37 & 45 & 12 \\ 
	16 & 0.61 & 33 & 50 & 12 \\
	17 & 0.75 & 47 & 49 & 12 
\enddata
\tablecomments{Slice numbers correspond to the numbers on Figure 7.}
\end{deluxetable}

\subsubsection{Results (Method1)}

The resulting family of $\tau$ versus $T_s$ curves for the analysis of Method 1 is shown in Figure 9.  The numbers next to each line correspond to the latitude slices shown in Figure 7 with measured variables listed in Table 1.  For Figure 9, we assume $p = 0.65$.  By adjusting $p$, the asymptotes of the $\tau$ versus $T_s$ curves change.  The spin temperature limits are proportional to p.  

By considering all latitude slices, an upper limit to the spin temperature of this HISA cloud ranges from 22 K to 46 K. Looking more closely at Figure 9, the upper limits to the spin temperature follow a pattern as we increase latitude.  On the bottom slice the upper limit to the spin temperature is at its maximum value of 46 K.  This decreases as we approach the center of the cloud to $\sim$ 22 K on slice \#12.  On higher latitude slices the upper limit to the spin temperature increases back to $\sim$ 40 K.  This is easily seen in Figure 10.  The upper limit spin temperature is plotted versus latitude for each slice.  At the lowest latitudes, the spin temperature is at its maximum and then quickly drops to its minimum around $b = 0^{\circ}$.  This corresponds to the center of the cloud seen in Figure 7.  It then slowly rises as we increase the latitude further.

The continuum brightness temperature, $T_{cont}$ consists of contributions from $T_{c,C}$ and $T_{c,D}$, the continuum located behind the cloud.  The numerical values for $T_{cont}$ in Table 1 consist of contributions from all four sources of diffuse continuum in Figure 5.  Therefore, in the above analysis, we assume that all of the continuum is behind the cloud.  Is this a valid assumption to make?  At this distance, as for the HI emission, a majority of the continuum must be from behind the cloud.  All of the marked HII regions in Figure 4 are located behind the cloud with typical distances of 1-3 kpc \citep{1987A&A...171..261C}.  If we have overestimated the background continuum, then we have also overestimated the upper limit spin temperature of the HISA cloud.  If only 75\% of the continuum is background, the cloud temperature decreases by at most 3 K.  In this case, the spin temperature varies from 19 K at the center of the cloud to 43 K at the edges, reaching a minimum at slice \#12, as before.  In the unlikely case that only 50\% of the continuum is background, the cloud becomes even cooler with center to edge spin temperatures ranging from 14 K to 40 K.

\subsubsection{Method 2:  Individual Continuum Sources}

We apply a second method to take advantage of several compact continuum sources in the vicinity of the HISA cloud.  Method 2 uses a slightly different geometry, illustrated in Figure 5, consisting of three lines of sight, `On', `Off', and `Src'.

$T_{src}$ is defined as the measured brightness temperature when the line of sight is towards a compact continuum source, illustrated by the `Src' line on Figure 5.  Similarly, $T_{on}$ is defined as the measured brightness temperature when the line of sight is slightly off the compact source, but still on the HISA cloud, seen as the `On' line.  $T_{cont}$ is the brightness temperature of the continuum and is different for the `Src' and `On' positions.   Since we do not have direct measurements of $T_{off}$, we interpolate the off-cloud brightness temperature across the absorption dip by fitting a parabola to the velocity profile.  The parabola is fitted to the data at velocities -9.1 to -3.3 km s$^{-1}$ and +1.6 to +4.9 km s$^{-1}$.

An example of the fitting is shown in Figure 11.  The solid line shows the velocity profile which includes the self-absorption dip, and the dashed line is our parabola fit to the off-line channels.  Other studies use different methods to find $T_{off}$.  These include fitting gaussian functions to the profile or using a straight line interpolation.  Due to complexities in the velocity profile at low latitudes \citep{1983AJ.....88..658H}, a gaussian fit may be no better than a parabola or straight line fit.  On the other hand, the straight line method underestimates the off-cloud brightness temperature.  We apply the parabola fit to every pixel in the cloud to find the off-cloud brightness temperature and create a background brightness temperature map at a velocity of -0.82 km s$^{-1}$, near the center of the feature.  This is shown in Figure 12 by two panels.  The upper panel is the HISA-free map where the HISA cloud in Figure 1 is removed, and the lower panel is the result of subtracting the original map in Figure 1 from the HISA-free map.  Both maps are shown at a velocity of -0.82 km s$^{-1}$.

This method requires a slight modification to Equation 6, because the `Src' and `On' lines of sight measure different values for $T_{cont}$, denoted as $T_{cont,src}$ and $T_{cont,on}$, respectively.  Figure 13 shows the positions for the `Src' and `On' lines of sight for continuum source \#1 in Table 2 with overlaid continuum contours.  The black cross marks the center of the continuum source, and the three white crosses mark the three points slightly off the continuum source, giving the `On' curves.  $T_{on}$ and $T_{off}$ are averaged for the three `On' points to give a single `On' curve which we compare to the `Src' curve. For the `Src' line of sight, Equation 6 becomes,

\begin{equation}
\tau = -ln(1 - \frac{T_{src} - T_{off}}{T_{s} - T_{cont,src} -pT_{off}})
\end{equation}
where $T_{cont,src}$ is the brightness temperature of the continuum source, most likely located behind the cloud.  For the `On' line of sight,

\begin{equation}
\tau = -ln(1 - \frac{T_{on} - T_{off}}{T_{s} - T_{cont,on} -pT_{off}})
\end{equation}

\noindent where $T_{on}$ replaces $T_{src}$ and $T_{cont,on}$ is the continuum brightness temperature at the `On' position.  Equations 7 and 8 each produce a $\tau$ vs. $T_s$ curve; one equation for the `Src' line of sight and a second equation for the `On' lines of sight. Because we are looking at the same portion of the HISA cloud, there is a solution for $\tau$ and $T_s$ that satisfies both measurements for a given $p$.  If the two curves cross, the result is 
 a unique optical depth and spin temperature. The physical separation between the `Src' and `On' positions is sufficiently small that the HISA cloud's spin temperature and opacity at the two locations are assumed to be the same.  In total, we analyze four continuum sources, listed in Table 2.

\begin{deluxetable}{ccccccc}
\tablewidth{0pc}
\tablecaption{Measured parameters for the four continuum sources analyzed in Method 2. }
\tablehead{
\colhead{Source\#}  & \colhead{Src;On} & \colhead{Latitude(Deg)} & \colhead{Longitude(deg)} & \colhead{$T_{src;on}$(K)} & \colhead{$T_{off}$(K)} & {$T_{cont,src;cont,on}$}}
\startdata
	1 & Src & 318.06 & -0.44 & -7 & 31 & 39 \\ 
	 & On & 318.20 & -0.41 & 30 & 71 & 14 \\ 
	 & On & 318.16 & -0.33 & 34 & 70 & 14 \\ 
	 & On & 317.99 & -0.43 & 29 & 68 & 15 \\ 
	2 & Sr
c & 318.10 & -0.01 & 16 & 37 & 28 \\ 
	 & On & 318.05 & -0.01 & 38 & 55 & 15 \\ 
	 & On & 318.11 & +0.04 & 41 & 59 & 15 \\ 
	 & On & 318.14 & -0.01 & 33 & 61 & 15 \\ 
	3 & Src & 317.89 & -0.01 & 15 & 35 & 26 \\
	 & On & 317.83 & -0.03 & 27 & 55 & 18 \\ 
	 & On & 317.90 & -0.11 & 27 & 57 & 17 \\
	 & On & 317.96 & -0.04 & 32 & 46 & 17 \\
	4 & Src & 317.86 & +0.16 & -15 & 11 & 51 \\
	 & On & 317.93 & +0.10 & 40 & 63 & 16 \\
	 & On & 317.92 & +0.16 & 47 & 68 & 16 \\ 
	 & On & 317.91 & +0.19 & 43 & 73 & 14
	 
\enddata
\end{deluxetable}

\subsubsection{Results (Method 2)}

 For the four continuum sources we analyze, we construct `Src' and `On' curves.  Figure 14 shows the `Src' curve as a solid line, and the averaged `On' curve as the dashed line for continuum source \#4, the strongest of the continuum sources. Dashed-dot lines represent column density curves, described later.

The three `On' curves are averaged together into one curve, the dashed line in Figure 14.  In our use of Equations 6, 7, and 8, there are three variables we wish to solve for:  $p$, $T_s$, and $\tau$.  By using the three `On' curves  and determining for which values of $p$ the `Src' curve lies between the two outer `On' curves, we obtain a range of acceptable values for $p$.  For continuum source \#4, $p$ ranges from 0.53 - 0.64. We find the upper limit spin temperature by averaging the `On' curves.  The value of $p$ is adjusted until the asymptote of the `Src' curve lines up with the asymptote of the average `On' curve.  This gives the upper limit to the spin temperature of the HISA cloud.  For source \#4, $T_s \leq 31$ $K$. The results are listed in Table 3.

\begin{deluxetable}{cccc}
\tablewidth{0pc}
\tablecaption{Measured spin temperatures and $p$ ranges for the four continuum sources.}
\tablehead{
\colhead{Source\#}  & \colhead{$p$} & \colhead{$T_{s,max}$(K)}  & \colhead{$T_{cont,src}$}}
\startdata
	1 & 0.59 -- 0.70 & 23 & 39 \\ 
	2 & 0.47 -- 0.83 & 29 & 28 \\ 
	3 & 0.26 -- 0.86 & 31 & 26 \\ 
	4 & 0.53 -- 0.64 & 31 & 51 
\enddata
\end{deluxetable}

A comparison of Method 1 and Method 2 is shown in Figures 10 and 15.  Figure 15 shows a plot of optical depth and column density (discussed in the following section) versus latitude.  The four solid squares represent the results found for the four continuum sources analyzed in Method 2.  The solid line with points illustrates the results for the constant latitude lines in Sections 3.1.1 and 3.1.2.  Aside from continuum source \#1, the values found in Method 2 all lie below the optical depth line in Method 1.  Error bars are included in Figure 15.

\subsection{Optical Depth and Column Density}

Looking at the $\tau$ vs. $T_s$ curves in Figure 9, the absolute lower limit optical depth corresponds to $T_s = 2.7$ K.  When the spin temperature is at its asymptotic value, $\sim$ 20-25 K at the center of the cloud, the optical depth is infinite.  Therefore we place constraints on the spin temperature to obtain a range of optical depths for the HISA cloud.  We choose to back off the upper limit spin temperature to 20 K, and set a range of 10-20 K to determine the optical depth.  We compute optical depths for all seventeen latitude slices  for spin temperatures of 10 K and 20 K with the results in Table 4.  Results for the continuum sources analyzed in Method 2 are provided in Table 5.  The change in optical depth resulting from $T_s$ changing from 10 K to  20 K is at most a factor of two.  Similar to the spin temperature, the optical depth follows a pattern as we move through each latitude slice in the cloud, assuming a constant $T_s$.  At the edges of the feature, the optical depth is at its minimum and reaches a maximum at slice \#12, near the center.   This is shown in Figure 15.  Each data point represents one of the seventeen latitude slices.  The upper and lower panels show optical depths and column densities for T$_s$ = 20 K and T$_s$ = 10 K, respectively.

We introduce column density curves, seen as the dash-dot lines in Figure 14.  The column density curves are given by

\begin{equation}
\tau(v) = \frac{N(v)}{CT_{s}\Delta{v}}
\end{equation}

\noindent where $N(v)$ is the column density in units of $cm^{-2}$, $\Delta$$v$ is the velocity width of the cloud, and C is $1.823\times 10^{18}$ $cm^{-2}$ $K^{-1}$ ($km$ $s^{-1})^{-1}$.  Curves with column densities of 1, 5, 10, and 20 times $10^{19}$ $cm^{-2}$ are drawn increasing towards the upper right in Figure 14. If the spin temperature and optical depth are known, the line of sight column density of the HISA cloud can be found.  Column densities are computed using input values from Tables 4 and 5.  Column densities for each latitude slice from Method 1 are given in Table 4, and column densities for continuum source positions from Method 2 are given in Table 5.  A plot of the column density versus latitude is shown in Figure 15.  As expected, the column density is lower at the edges of the cloud and increases to a maximum value at slice \#12. Due to slight changes in optical depth at differing spin temperatures, the peak column density occurs at different slices in Figure 15.

\begin{deluxetable}{ccccc}
\tablewidth{0pc}
\tablecaption{Measured optical depths and column densities for seventeen latitude slices.}
\tablehead{
\colhead{Latitude Slice \#} & \colhead{$\tau$(10 K) } & \colhead{$\tau$(20 K)}  & \colhead{$N_H$($10^{19}$cm$^{-2}$, $T_s$ = 10 K) } & \colhead{$N_H$($10^{19}$cm$^{-2}$, $T_s$ = 20 K) }}
\startdata 

	1 & 0.25$\pm$0.07  & 0.33$\pm$0.10 & 1.1$\pm$0.31 & 2.9$\pm$0.90 \\
	2 & 0.37$\pm$0.07 & 0.49$\pm$0.09 & 1.7$\pm$0.31 & 4.4$\pm$0.81 \\
	3 & 0.44$\pm$0.07 & 0.56$\pm$0.10 & 2.0$\pm$0.31 & 5.0$\pm$0.90 \\
	4 & 0.80$\pm$0.09 & 1.2$\pm$0.15  & 3.6$\pm$0.40 & 11$\pm$1.3 \\
	5 & 0.76$\pm$0.08 & 1.1$\pm$0.13  & 3.4$\pm$0.36 & 10$\pm$1.2 \\
	6 & 0.83$\pm$0.08 & 1.3$\pm$0.15  & 3.8$\pm$0.36 & 11$\pm$1.3 \\
	7 & 1.0$\pm$0.10 & 1.6$\pm$0.19  & 4.5$\pm$0.45 & 14$\pm$1.7\\
	8 & 1.2$\pm$0.11 & 2.0$\pm$0.24   & 5.4$\pm$0.49 & 18$\pm$2.2 \\
	9 & 0.99$\pm$0.10 & 1.6$\pm$0.18 & 4.5$\pm$0.45 & 14$\pm$1.6 \\
	10 & 1.1$\pm$0.11 & 1.8$\pm$0.24  & 4.8$\pm$0.49 & 17$\pm$2.2 \\
	11 & 1.0$\pm$0.11 & 1.8$\pm$0.23  & 4.7$\pm$0.49 & 16$\pm$2.1 \\
	12 & 1.1$\pm$0.13 & 2.3$\pm$0.40  & 5.1$\pm$0.58 & 21$\pm$3.6 \\
	13 & 0.99$\pm$0.13 & 2.0$\pm$0.37 & 4.5$\pm$0.58 & 18$\pm$3.3 \\
	14 & 0.48$\pm$0.12 & 0.83$\pm$0.20 & 2.2$\pm$0.54 & 7.5$\pm$1.8 \\
	15 & 0.30$\pm$0.11 & 0.47$\pm$0.17 & 1.3$\pm$0.49 & 4.3$\pm$1.5 \\
	16 & 0.68$\pm$0.12 & 1.2$\pm$0.22  & 3.1$\pm$0.54 & 11$\pm$2.0 \\
	17 & 0.06$\pm$0.09 & 0.09$\pm$0.13 & 0.28$\pm$0.40 & 0.78$\pm$1.2

\enddata
\end{deluxetable}

\begin{deluxetable}{ccccc}
\tablewidth{0pc}
\tablecaption{Measured optical depths and column densities for the four continuum sources in Method 2.}
\tablehead{
\colhead{Source \#} & \colhead{$\tau$(10 K) } & \colhead{$\tau$(20 K)}  & \colhead{$N_H$($10^{19}$cm$^{-2}$, $T_s$ = 10 K) } & \colhead{$N_H$($10^{19}$cm$^{-2}$, $T_s$ = 20 K) }}
\startdata 
	1 & 1.4$\pm$0.12 & 3.1$\pm$0.61 & 6.5$\pm$0.54 & 28$\pm$5.5 \\
	2 & 0.73$\pm$0.10 & 1.2$\pm$0.18 & 3.3$\pm$0.45 & 10$\pm$1.6 \\
	3 & 0.66$\pm$0.09 & 1.0$\pm$0.14 & 3.0$\pm$0.40 & 9.0$\pm$1.3\\
	4 & 0.79$\pm$0.33 & 1.2$\pm$1.1 & 3.6$\pm$1.5 & 11$\pm$9.9
\enddata
\end{deluxetable}

\subsection{Errors}
When analyzing this cloud, errors were introduced in the measurement of the `On' and `Off' brightness temperatures.  Other errors include fitting the offline emission, determining the velocity of the cloud, and determining a value for `p'.  In this section we attempt to account for these.

 In measuring $T_{off}$ we interpolate the brightness temperature over the self-absorption dip in Figure 11.  A conservative estimate of the uncertainty in $T_{off}$ is 3 K.  The uncertainties in $T_{src}$, $T_{cont,src}$, and $T_{cont,on}$ are insignificant because we obtain these without averaging or interpolating.  For the `On' curve, the uncertainty in $T_{off}$ remains, but there is added uncertainty in $T_{on}$.  This uncertainty is also taken to be 3 K.  The standard procedure for propagation of errors is used, and the uncertainties in $\tau$, for a specific $p$, are given by:

\begin{equation}
\sigma_{\tau_{src}} = \mid\sigma_{T_off}\frac{T_{cont,src} + pT_{src} - T_s}{(T_s - T_{cont,src} - pT_{off} - T_{src} +T_{off})(T_s - T_{cont,src} - pT_{off})}\mid \sim 0.10
\end{equation}

Typical optical depth errors for a spin temperature of 10 K are 0.10.  These are shown as the vertical error bars in Figure 14.  The uncertainties in $T_s$ are found by propagation of errors and are denoted by the horizontal error bars in Figure 14.  The uncertainties in the spin temperature are:

\begin{equation}
\sigma_{T_{s,src}} = \mid\sigma_{T_{off}}{(p - \frac{1}{1 - e^{-\tau}})}\mid \sim 3 K
\end{equation}

Here we have used p = 0.65 and an optical depth of 1 as an example.  The subscript `Src' refers to uncertainites in the `Src' curve.  Uncertainties associated with the `On' curve will be of the same form.  For illustration, two points are plotted with uncertainties in $\tau$ and one point is plotted with uncertainties in $T_s$ for both curves.  Seen in Figure 14, the error bars in the `On' curves are larger than those in the `Src' curves.  This is expected due to the uncertainty in measuring $T_{on}$ which we do not have in measuring $T_{src}$.

If we look at the limiting case when $T_s = 0$ K and there is no continuum, Equation 10 reduces to:

\begin{equation}
\sigma_{\tau_{src}} = \mid\sigma_{T_{off}}(\frac{T_{src}}{T_{off}(T_{src} + (p-1)T_{off})})\mid
\end{equation}

\noindent The $p - 1$ term is a negative value unless $p = 1$, and the dominant source of error is in the $T_{src}$ and $T_{off}$ terms.  In the case of the `Src' curve, the error in $T_{off}$ will dominate.  If $T_{src}$ $\sim$ $T_{off}$, then $\sigma_{\tau}$ $\rightarrow$ $\frac{\sigma_{T_{off}}}{T_{off}}$.  In considering the limit where $\tau$ $\rightarrow$ $\infty$, the uncertainty in $T_s$ is given by $p \sigma_{T_{off}}$, and we are limited by our ability to measure $T_{off}$.  Uncertainties in optical depth and column density are given in Tables 4 and 5.

\section{A Second HISA Feature}

We apply Method 2 to another HISA cloud nearby.  The second cloud is centered on $l = 319^{\circ}$ and  $b = 0.4^{\circ}$ at the same velocity, v = -1.1 km s$^{-1}$, seen in Figure 16.  Fortunately there is a compact source with brightness temperature of 46 K just behind the middle of the cloud.  Using the average of three `On' positions, the upper limit to the spin temperature is 20 K, and the range of $p$ values is 0.47 - 0.66.  We calculate a range of optical depths and column densities for spin temperatures, $T_s = 10 - 15 K$. The optical depth ranges from 1.6 to 2.1 and the column density ranges from $2 \times 10^{20}$ $cm^{-2}$ to $4 \times 10^{20}$ $cm^{-2}$.  These are similar to the results for the first HISA cloud. With such similarities in spin temperature, $p$, optical depth, column density, velocity, and spatial postion, we conclude the two clouds are not two separate clouds, but clumps of gas that are parts of a single, larger complex.  A profile through the second cloud is shown in the lower panel of Figure 16.

\section{Discussion}

The coldest and most dense of the diffuse, atomic clouds are those that stand out in HI surveys due to absorption of the background $\lambda$21cm emission.  By determining the physical parameters of this HI cloud to a reasonable accuracy, we can use it as an example to compare the conditions inside HI clouds selected on the basis of their HISA signature, to those in the more extensively studied molecular clouds, selected by their CO emission.  In Section 3 we show that this HISA cloud has an upper limit spin temperature, $T_s$ $\sim$ 25 K, corresponding to an infinite optical depth.  At a spin temperature of 20 K the column density is, $N_H$ $\sim$ $2\times10^{20}$ $cm^{-2}$, and the optical depth is, $\tau$ $\sim$ 2.  If the spin temperature of the cloud is lower, then the column density and optical depth of the cloud will also be lower.  For a spin temperature of 10 K, the column density and optical depth are, respectively, $N_H$ $\sim$ $5\times10^{19}$ $cm^{-2}$ and $\tau$ $\sim$ 1.  These are similar to the values of the HISA clouds of \citet{2000ApJ...540..851G} and the core of the giant HI cloud of \citet{2001ApJ...555..868M} and are typical of clouds detected from inspection of low latitude HI emission surveys without pre-selection on the basis of other cloud tracers.  The alternative approach to searching for HISA by
starting from a sample of molecular clouds, as 
taken by \citet{1974AJ.....79..527K}, \citet{1980ApJ...242..416L}, \citet{1984A&A...140..303L}, and \citet{Li}, generally finds clouds with lower
optical depths that are taken to have even colder
temperatures (similar to the molecular cloud temperatures of 10
to 15 K) and HI column densities typically smaller
than the value we find here.  This may well be a 
selection bias; the difference in physical characteristics
may result from the way of choosing the sample of 
objects.  The point is that there is a great deal
of self-absorption in the HI surveys, much of it
imperceptible.  The clouds that cause this 
HISA phenomenon may span a critical range in density,
temperature, and column density values that fills in
the ``missing link'' between the better studied dense
molecular clouds (with densities above 10$^3$ cm$^{-3}$
so that the $\lambda$2.7 mm lines of CO are thermalized)
and the diffuse atomic clouds seen in absorption toward
continuum sources, that have temperatures of 25 to 100 K
and densities of a few tens to 100 cm$^{-3}$ or so
\citep{1988gera.book...95K, 1990ARA&A..28..215D, Heiles}.

To determine if our HISA cloud is associated with molecular gas, we overlay our 21cm maps with contours taken from corresponding channel maps of $^{12}$CO emission from \citet{1987ApJ...322..706D, 2001ApJ...547..792D} in Figure 17.  Although the CO velocity resolution, $\delta$$v$ = 1.30 $km$ $s^{-1}$, is coarse, there is a clear spatial correlation between CO and HI.  The peak CO brightness temperature at the center of the first HISA cloud is about 0.3 K.  Other CO clouds are seen with stronger peaks, as well as a large complex at b = -4\arcdeg, G317-4, that \citet{2001ApJ...547..792D} show having a peak CO brightness temperature of 4 K.  The
CO features on Figure 17 may well be associated with
that minor GMC (giant molecular cloud) complex.  But why
do the other, brighter CO clumps not show any obvious
self-absorption in the 21-cm maps?  In Figure 17, at l = 316.5$^\circ$ and l = 319.4$^\circ$ there are two CO features not showing any signs of HI self-absorption.  Instead, they appear to be associated with bright HI emission.  This is indeed a real feature.  A closer look at the $^{12}$CO velocity profile shows the CO brightness temperature peaks with the HI.  It is possible that geometry of the HI gas may be such that the HISA gas is unobservable, but there is no reason that this should be the case.  The HI emission does not vary dramatically across the map in Figure 12, so it is unlikely that the geometry of these clumps is different from that of the HISA cloud.

In the interior of a dense molecular cloud, where no
photo-dissociating uv radiation can penetrate, there
is a small residual density of atomic hydrogen due to
cosmic ray ionization \citep{1971ApJ...165...41S, Li} that gives an HI density of about 2 cm$^{-3}$ independent of the molecular hydrogen density.  Dividing the column density (2 $\times$ 10$^{20}$ cm$^{-2}$) by this results in a depth of $\sim$ 30 pc.  For a spin temperature of 20 K, the depth is 40 pc.  These are 4-5 times higher than the diameter of our cloud in the plane of the sky ($\lesssim$ 8 pc). However, this density is an absolute minimum and assumes no photodissociating radiation is penetrating the cloud.  With our column densities, it is likely that some radiation is penetrating the cloud.  In that case, $n_H$ will be larger.  To obtain a line of sight depth of 8 pc, the mean HI density needs to be $\sim$ 10 cm$^{-3}$.  The result is a thermal pressure of $\sim$ 200 K cm$^{-3}$, an order of magnitude below the mean value of the local ISM, suggesting that the remaining density needed to account for the thermal pressure resides in the cloud as molecular gas.  This is in agreement with the presence of $^{12}$CO emission at the position of the HISA cloud.  Our HI densities are similar to the prediction of photo-dissociation
models \citep{1993ApJ...402..195W} for the outer layers of a
GMC, such as GRSMC 45.6+0.3 studied by \citet{2002ApJ...566L..81J}.
But our clouds are not the envelope of a GMC, they are small
CO clumps, and their larger CO clump neighbors do not show
similar HI self-absorption.  Also unlike GRSMC 45.6+0.3 and
other GMC-related HISA clouds, there is no obvious
obscuration visible on the ESO sky survey Schmidt plates
corresponding to the position of our HISA clouds.

To satisfy the thermal pressure constraints of the ISM, the HI density fraction is $\sim$10\% and the HISA cloud is composed primarily of molecular gas.  However, the CO contours on Figure 17 show that the $^{12}$CO emission is not particulary strong, compared to the other two CO clumps in the figure.  The peak brightness temperature of the CO clump associated with the HISA cloud is only 0.3K, while the temperatures of the other two clumps are 0.4K and 0.8K.  Can this cloud be a molecular cloud, as suggested by the atomic and molecular densities from above?   Recently, \citep{Heiles} have shown that HI clouds are not isotropic, but are instead blobby sheets with length-to-thickness ratios ranging from 100-300.  If the length-to-thickness ratio of this cloud is 100 and we are seeing the cloud face on, the HI density needed for the column densities found in Section 3.2 is $\sim$ 200 cm$^{-3}$.  This gives a thermal pressure similar to that of the local ISM and the HI fraction is now much larger, such that the cloud is dominated by atomic gas.

One possible scenario for the physical conditions in this cloud,
and perhaps in other clouds that stand out as HISA in HI surveys
but not in molecular emission or optical obscuration, is that 
they are in transition between a primarily atomic and a primarily
molecular state.  The time scale for equilibrium to be achieved
in molecule formation and destruction is typically $<$10$^6$ years,
so a cloud can make a fairly rapid transition between mostly
molecular to mostly atomic if the external conditions change.
The change in either the external radiation field or the ambient
pressure that can force this transition can be small \citep{1993ApJ...411..170E}.  If this cloud recently separated from the
vicinity of G317-4 and entered an environment with lower pressure
or higher radiation field, it may have quite recently been 
through a large scale process of photodissociation.  Alternatively,
it may be a hold-out, the last of the clouds in its neighborhood
to make the transition from atomic to molecular.  The atomic to molecular transition was proposed for the giant HI cloud observed by \citet{2001ApJ...555..868M}.  Either way,
this cloud is an enigma; more examples are needed to help
us understand the role of such clouds in the atomic-molecular
transition.

\section{Conclusions}

We have presented two methods of analyzing a neutral hydrogen self-absorption cloud in the Southern Galactic Plane Survey using a four component radiative transfer model.  After measuring the on-cloud, off-cloud, and continuum brightness temperatures, we use both models to find $p$, the fraction of HI emission originating behind the cloud, $T_s$, the spin temperature, and $\tau$, the optical depth.  Our analysis shows that these clouds contain very cold gas with spin temperatures $\le$ 20 K, possibly reaching as low as 10 K.  We find a correlation with $^{12}$CO emission, but other CO clumps are not associated with 21cm self-absorption.

The cloud presented in this paper stands out in our initial search for self-absorption in the SGPS.  There may be few clouds as dramatic as this, but it is likely that smaller, less spectacular clouds are numerous.  Other high resolution HI studies, such as the Canadian Galactic Plane Survey \citep{2000ApJ...540..851G}, reveal a number of small self-absorption clouds.  To avoid the bias of finding only the largest and most pronounced features, an objective, automated  search techinque is needed.  To develop a better understanding of the cold neutral medium, we intend to search the entire SGPS.

\bibliography{references2}

\begin{figure}
\epsscale{1.0}
\plotone{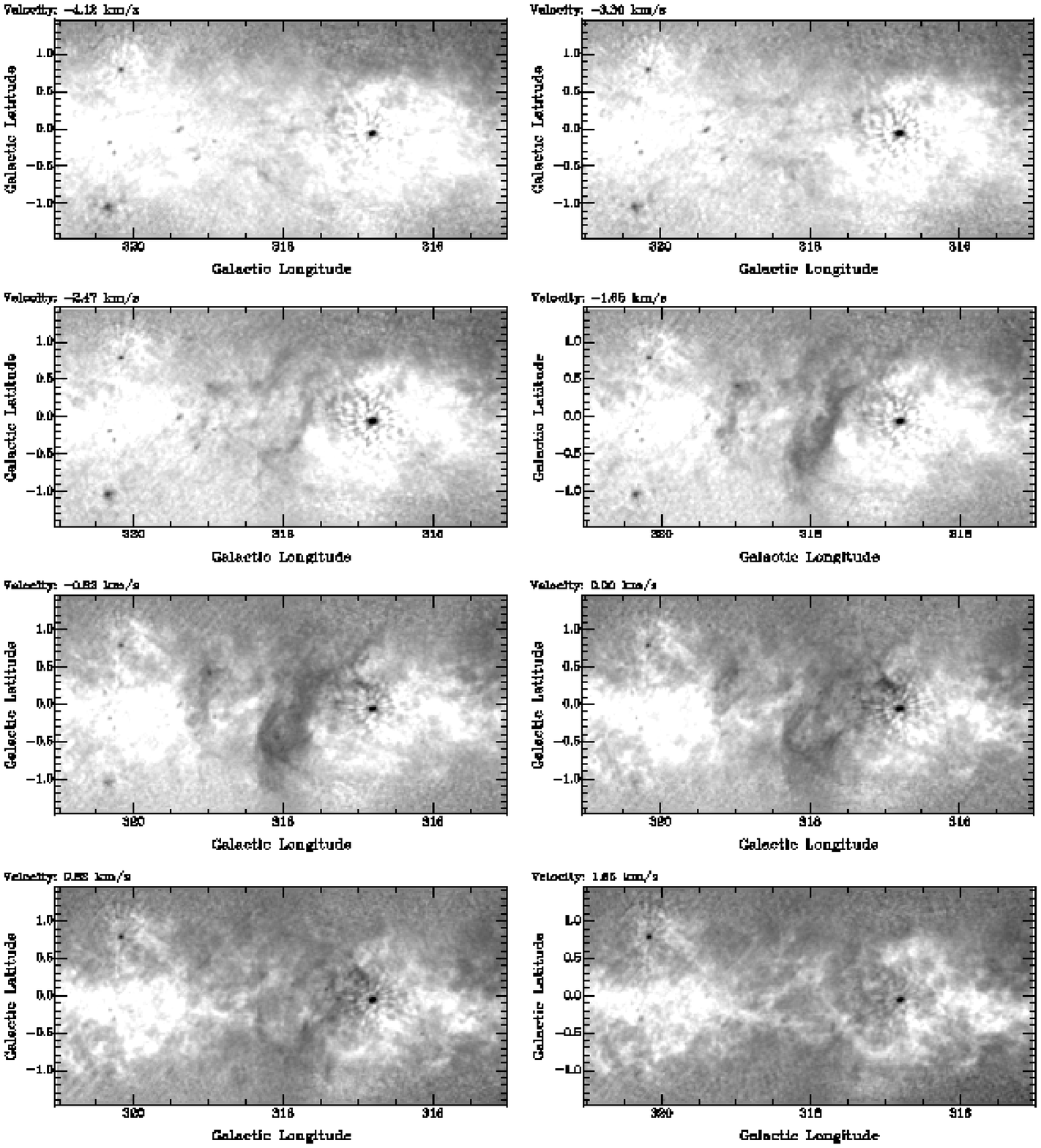}
\caption{Eight images of an HI self-absorption cloud at coordinates l = 318.0$^{\circ}$ and b = -0.5$^{\circ}$ with velocities ranging from -4.1 $km$ $s^{-1}$ to +1.7 $km$ $s^{-1}$.  The gray-scale brightness temperature ranges from 0 K(black) to 100 K(white).  }
\end{figure}

\begin{figure}
\epsscale{0.6}
\plotone{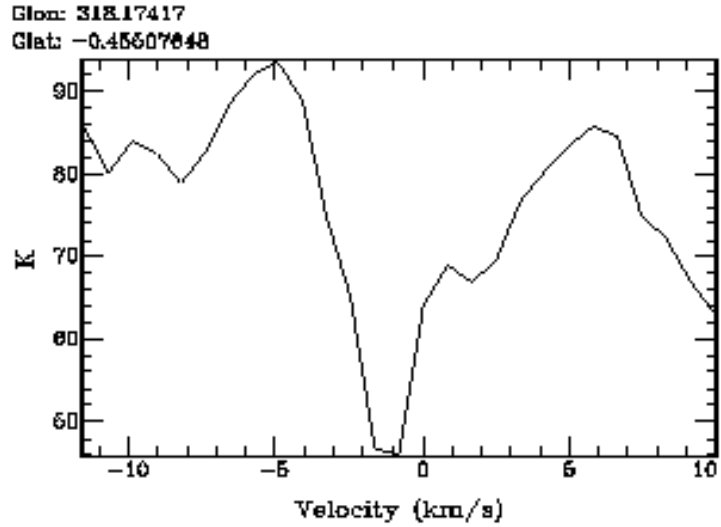}
\caption{A velocity profile of the cloud at coordinates $l = 318.2^{\circ}$ and $b = -0.46^{\circ}$.  The steepness of the dip and narrow width of the line are evidence that this is a self-absorption cloud.}
\end{figure}

\begin{figure}
\epsscale{0.6}
\plotone{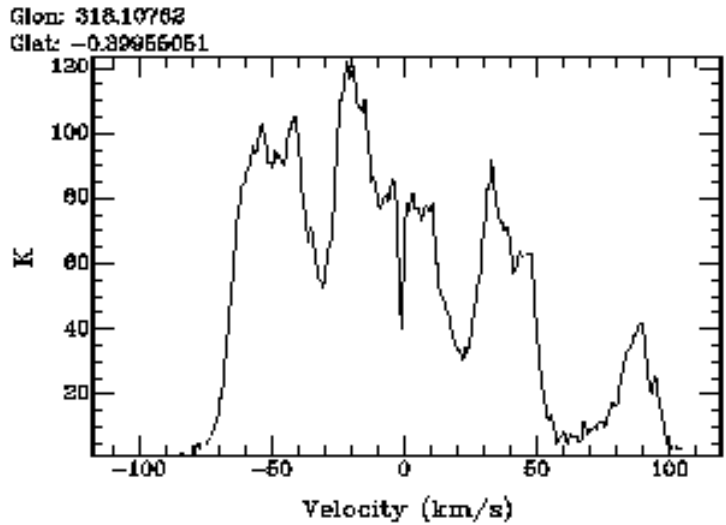}
\caption{A second velocity profile with a broader range in velocities.  The HISA cloud is represented by the sharp dip in the profile near v = 0 $km$ $s^{-1}$.}
\end{figure}

\begin{figure}
\epsscale{0.6}
\plotone{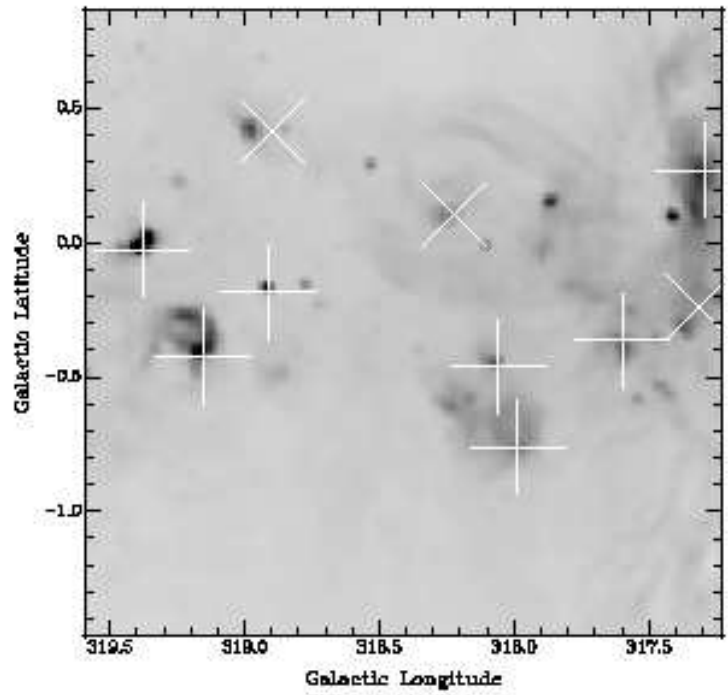}
\caption{21cm continuum image of the region near the HISA cloud.  Catalogued HII regions and SNR's are labled by +'s and X's, respectively.}
\end{figure}

\begin{figure}
\epsscale{1.0}
\plotone{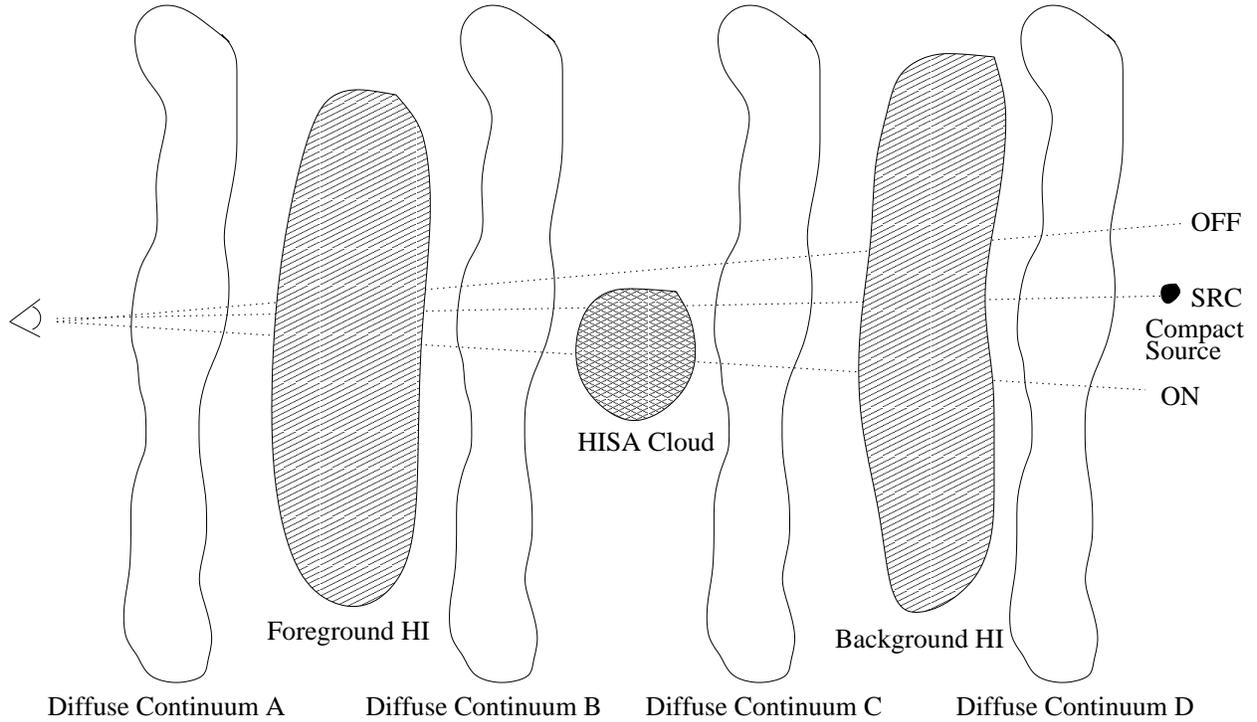}
\caption{An illustration of the four component model used in both methods.  In Method 1, only the `On' and `Off' lines of sight are used.  These two lines illustrate the geometry used to find $T_{on}$ and $T_{off}$.  The `Src' line of sight and the compact source are not used.  Method 2 takes advantage of the `Src' line of sight to find $T_{Src}$ and the discrete continuum source.} 
\end{figure}

\begin{figure}
\epsscale{0.6}
\plotone{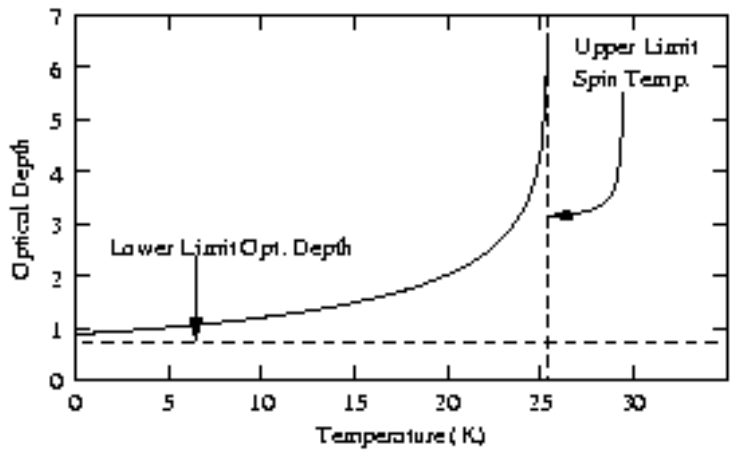}
\caption{An example $\tau$ vs. $T_s$ curve.  The upper limit spin temperature and lower limit optical depth values are labeled.  For this curve, $p = 0.65$, and the inputs for slice \#8 on Figure 7 and Table 1 are used.}
\end{figure}

\begin{figure}
\epsscale{0.6}
\plotone{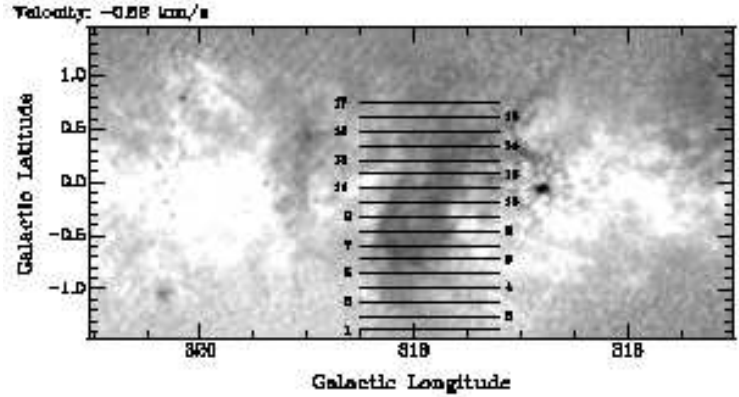}
\caption{An image of the region containing the HISA cloud.  The cloud is centered on coordinates $l = 318^{\circ}$ and $b = -0.5^{\circ}$.  The latitude slices from Method 1 are labeled accordingly.  The brightness temperature scale ranges from 0K(black) to 100K(white).}
\end{figure}

\begin{figure}
\epsscale{0.6}
\plotone{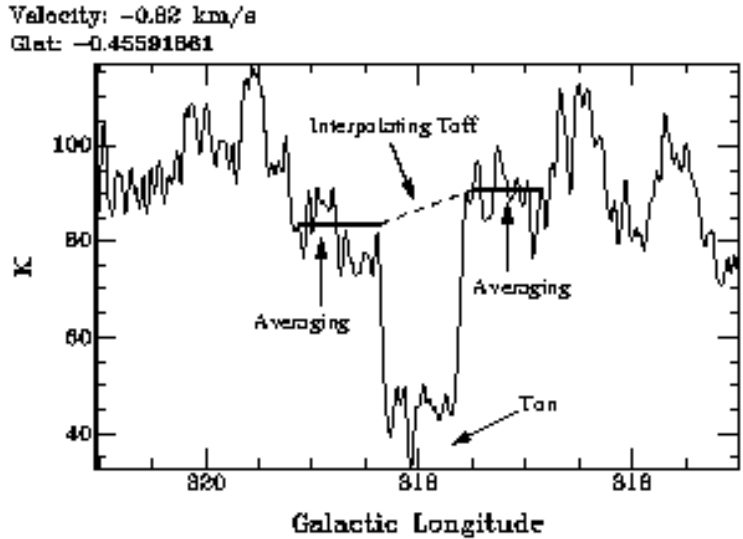}
\caption{A brightness temperature-longitude profile for slice \#8.  The straight solid lines show the portions of the profile that are averaged, and the dotted line shows the region across the absorption dip where $T_{off}$ is interpolated.  After interpolating, the continuum brightness temperature is subtracted to get the $T_{off}$ values in Table 1.}
\end{figure}

\begin{figure}
\epsscale{0.6}
\plotone{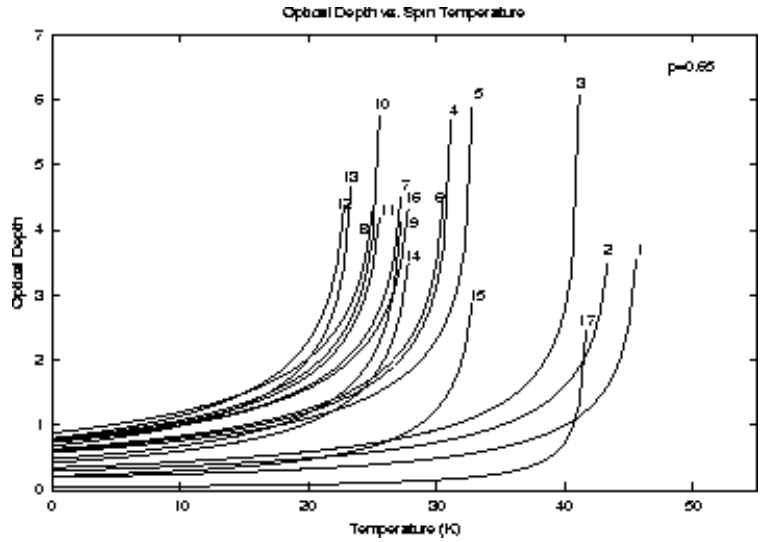}
\caption{Each line represents a $\tau$ vs. $T_s$ curve for one latitude slice, and the numbers correspond to the latitude slices on Figure 7.  For these curves, $p = 0.65$.}
\end{figure}

\begin{figure}
\epsscale{0.6}
\plotone{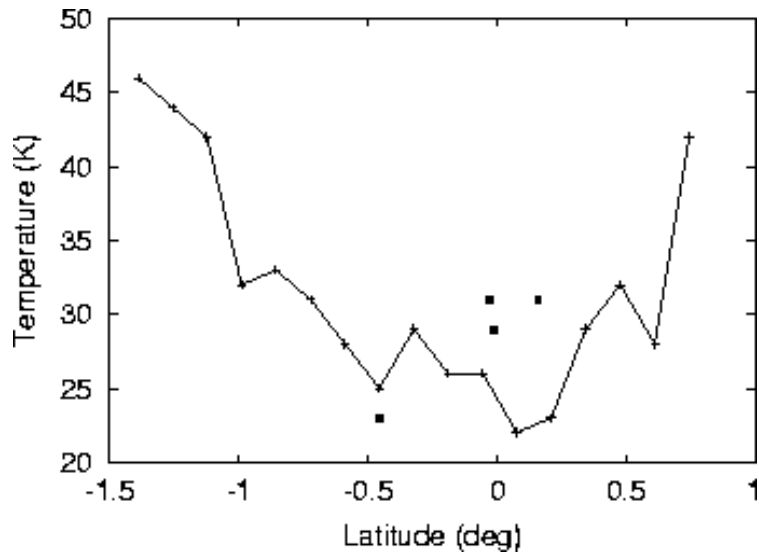}
\caption{A plot of the upper limit spin temperature versus latitude for each slice shown on Figure 7.  Each point shows the asymptotic value of the spin temperature of a single curve on Figure 9.  The filled squares represent upper limit spin temperatures found using four continuum sources in Section 3.1.3.
}
\end{figure}

\begin{figure}
\epsscale{0.6}
\plotone{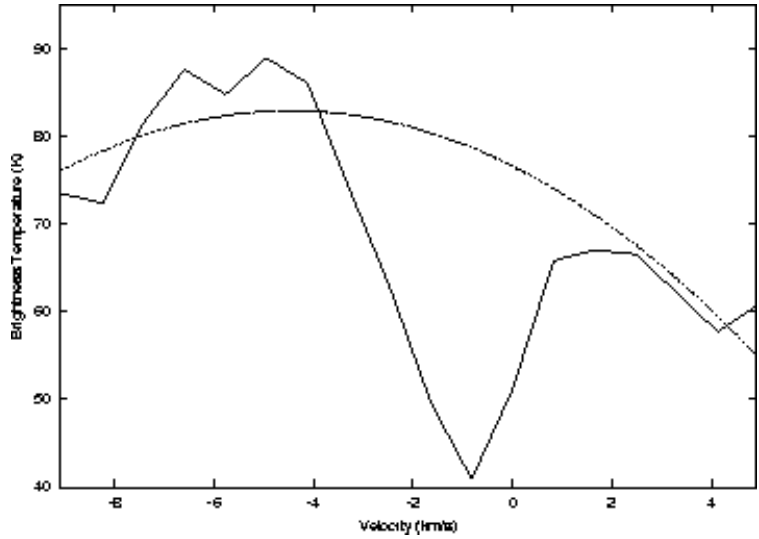}
\caption{The velocity profile at coordinates $l = 318.1^{\circ}$, $b = -0.47^{\circ}$.  The solid line is the profile and the dashed line is our parabola fit to the data.  Only the velocity channels -9.1 to -3.3 km s$^{-1}$ and +1.6 to +4.9 km s$^{-1}$ are fit.}
\end{figure}

\begin{figure}
\epsscale{1.2}
\plottwo{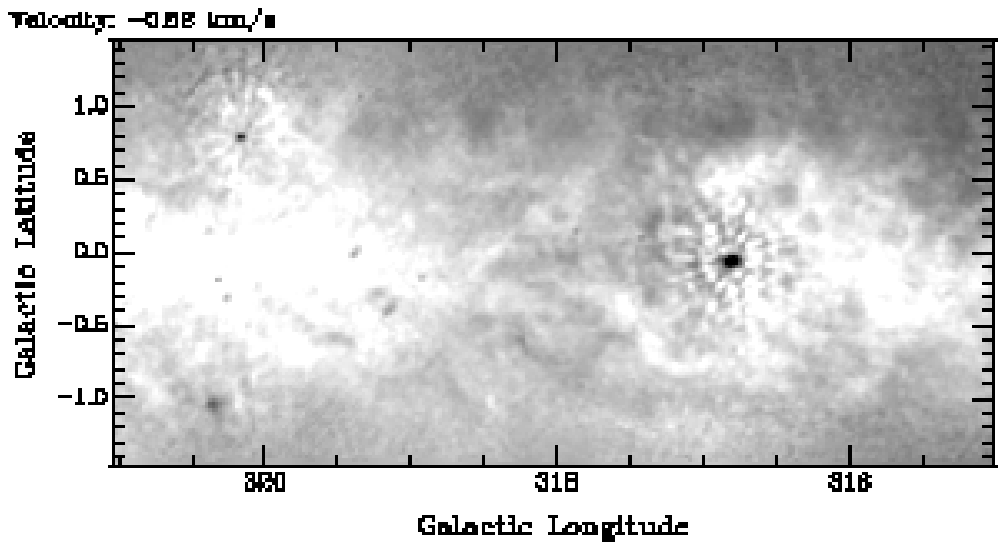}{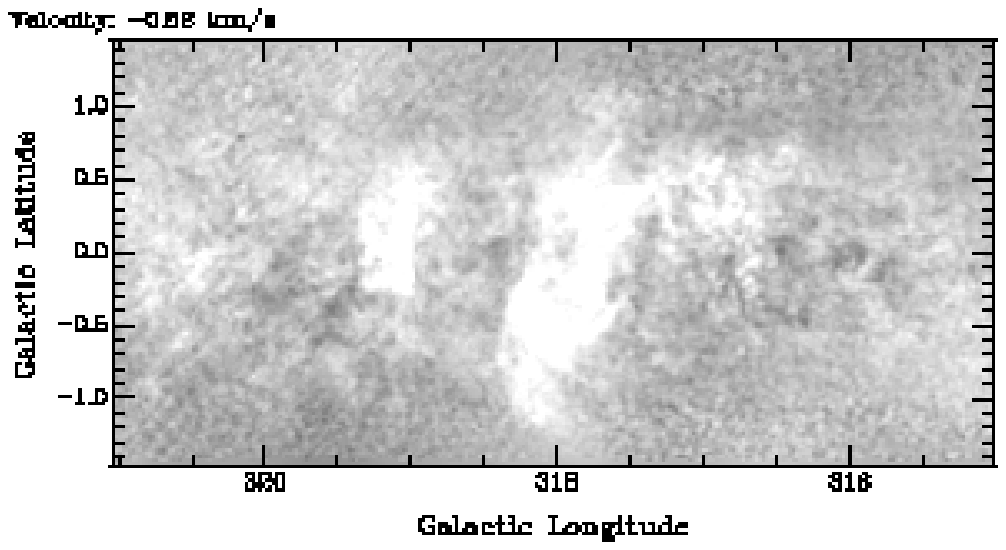}
\caption{Two HI images at a velocity of -0.82 km s$^{-1}$.  The upper panel shows the HI emission with the HISA cloud removed from the original v = -0.82 km s$^{-1}$ image in Figure 1.  In the lower panel, the original image in Figure 1 has been subtracted from the image in the upper panel.  The HISA cloud can clearly be seen as the bright white regions.  The brightness temperature in the upper panel ranges from 0(Black) to 100 K(White) and -75(Black) to 25 K(White) in the lower panel.}
\end{figure}

\begin{figure}
\epsscale{0.6}
\plotone{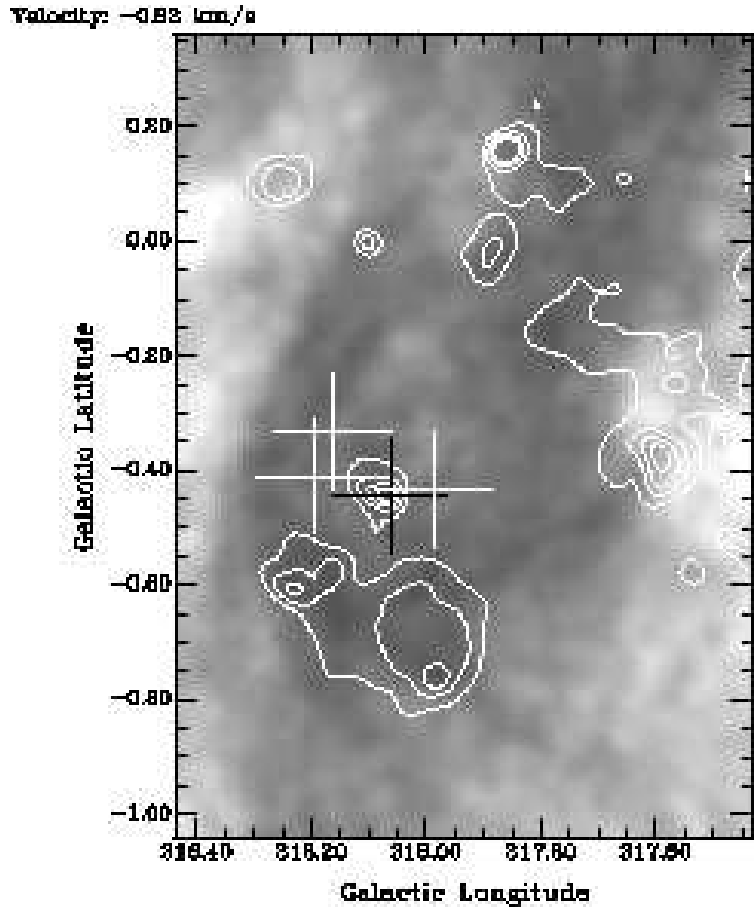}
\caption{A close-up of the HISA cloud in Figure 1 with overlaid continuum contours.  The contour levels are 20, 25, 30, and 35 K.  The black cross centered on the continuum source marks the position where $T_{src}$ is measured.  The three white crosses mark the three reference positions where $T_{on}$ is measured.  The gray-scale ranges from 0 K(Black) to 100 K(White).}
\end{figure}

\begin{figure}
\epsscale{0.7}
\plotone{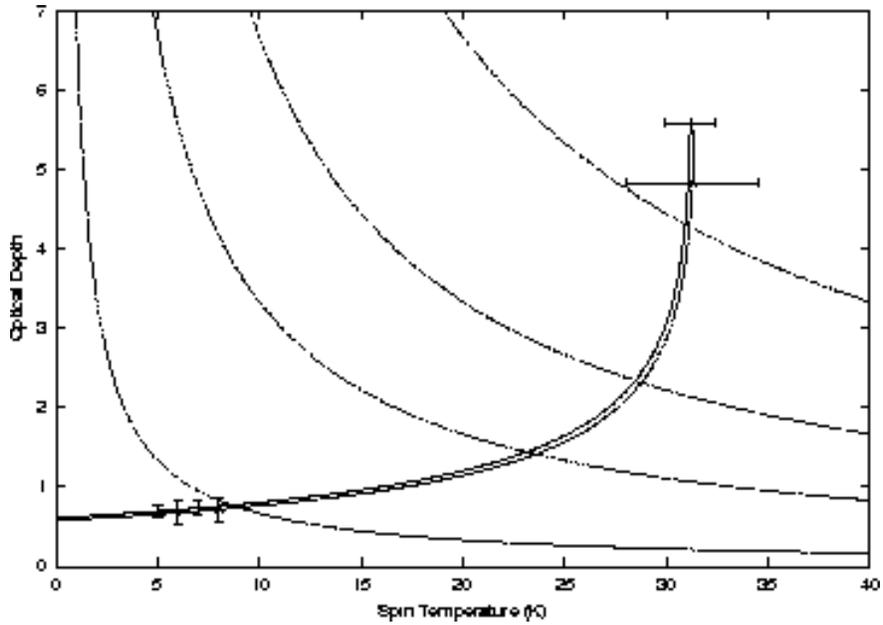}
\caption{The results of Method 2 for continuum source \#4.  The `On' curve is shown as a dashed line, and the `Src' curve is shown as a solid line.  The dash-dotted lines are column density lines of $N_H = $ 1, 5, 10, and 20 times $10^{19}$ $cm^{-2}$, increasing as optical depth and spin temperature increase.  Error bars are included, and for this source, $ p = 0.60$.}
\end{figure}

\begin{figure}
\epsscale{1.5}
\plottwo{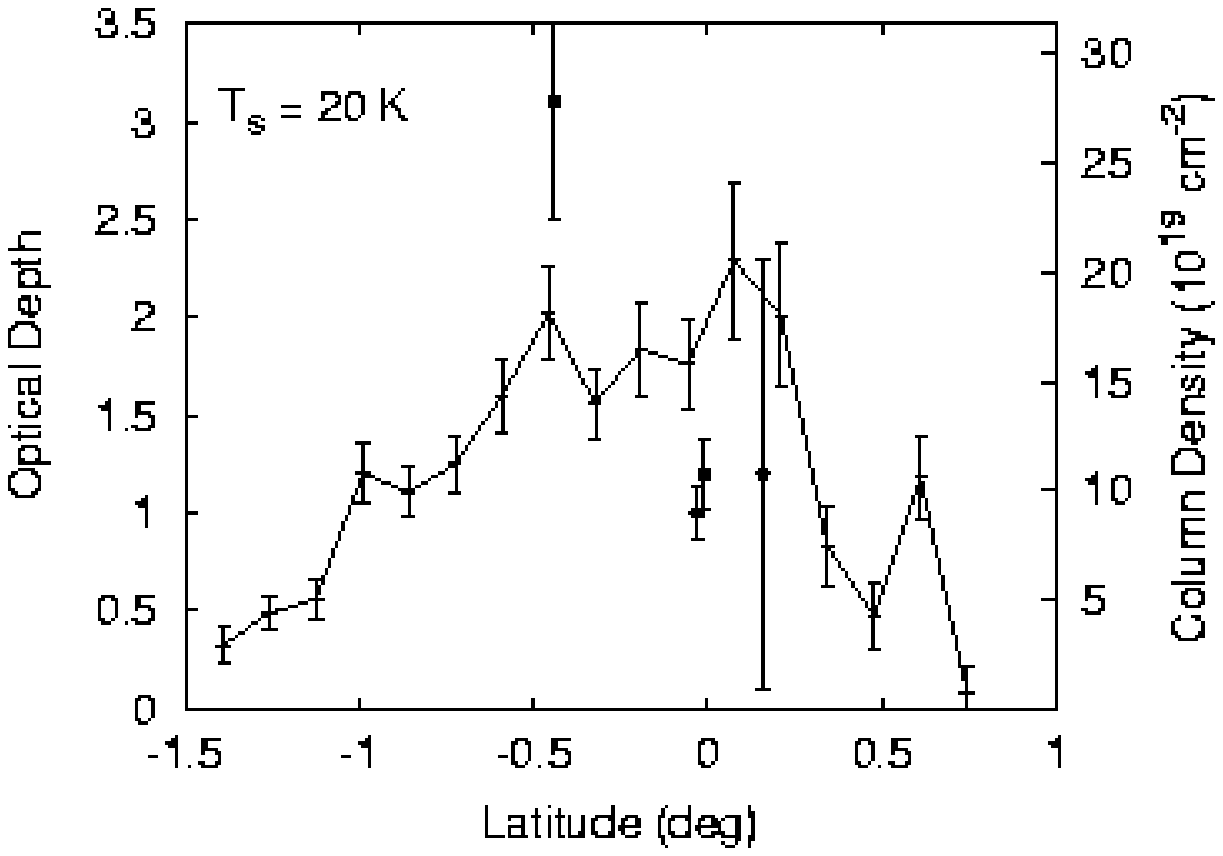}{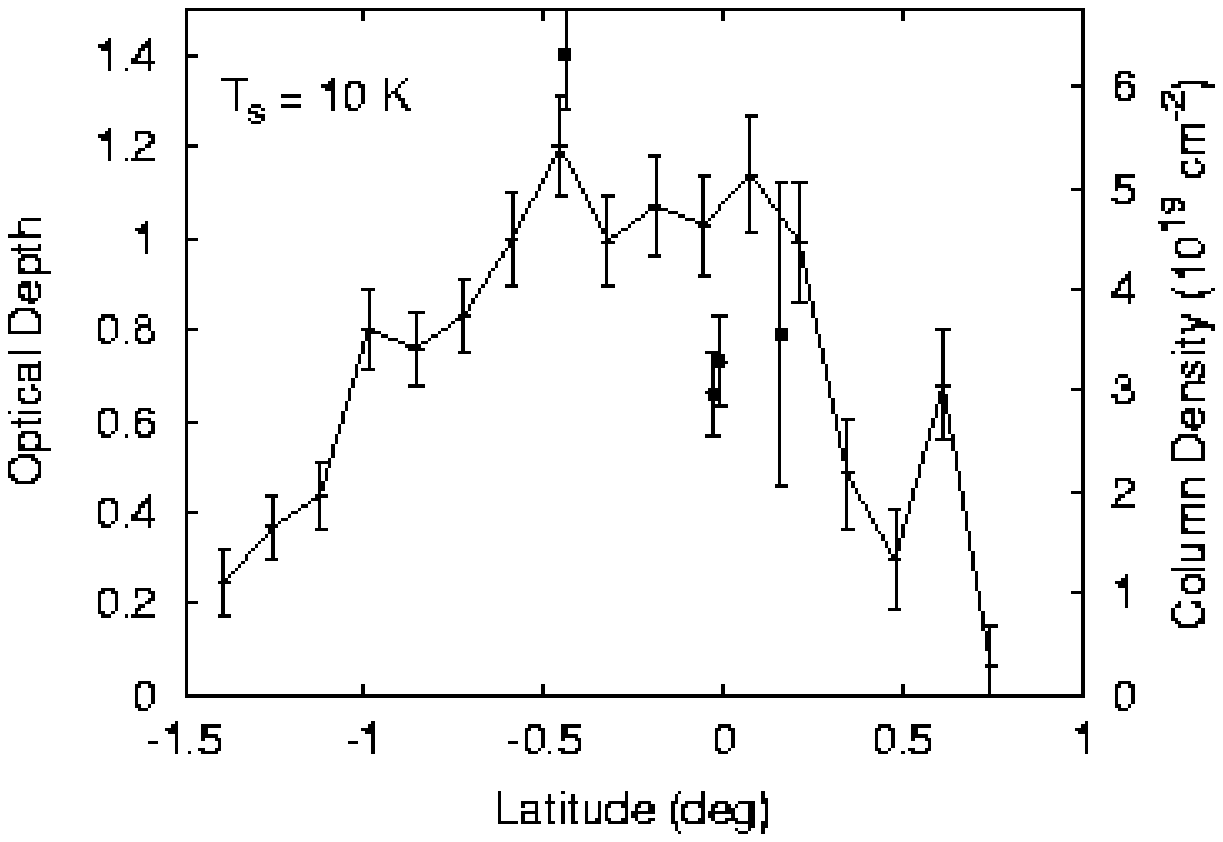}
\caption{Two plots of optical depth and column density versus latitude.  Each point represents a latitude slice from Figure 7.  The optical depth and column density for $T_s$ = 20 K are on the top panel and the corresponding values for $T_s$ = 10 K are plotted on the lower panel.  Filled squares represent optical depths for the four continuum sources analyzed in Method 2.}
\end{figure} 

\begin{figure}
\epsscale{1.4}
\plottwo{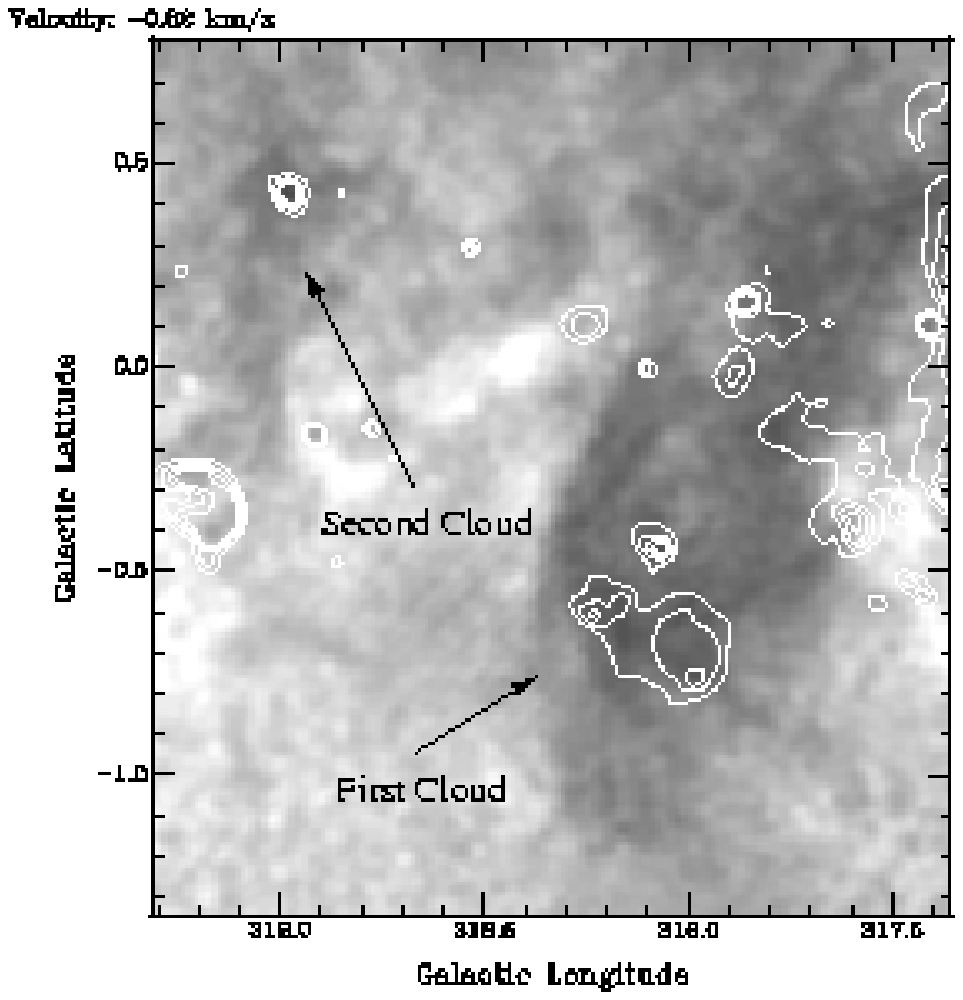}{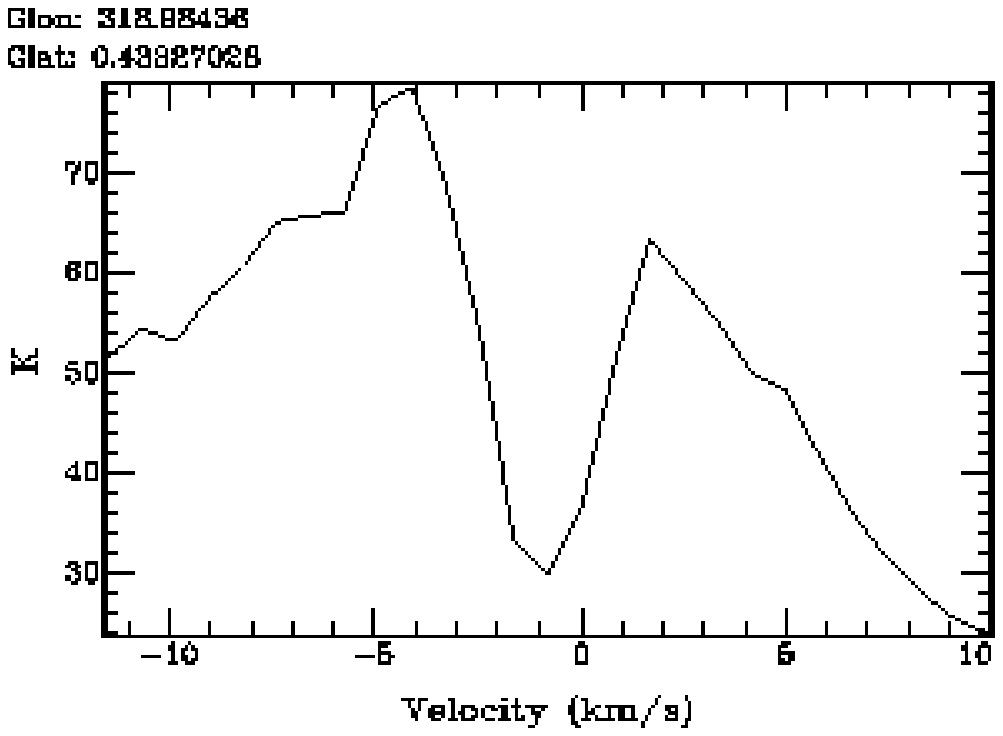}
\caption{The upper panel is an HI image of our HISA cloud that includes a second cloud at l = 319${^\circ}$ and b = 0.4$^{\circ}$.  The lower panel shows the velocity profile centered on the second feature. The gray scale ranges from 0K(Black) to 100K(White).}
\end{figure}

\begin{figure}
\epsscale{1.0}
\plotone{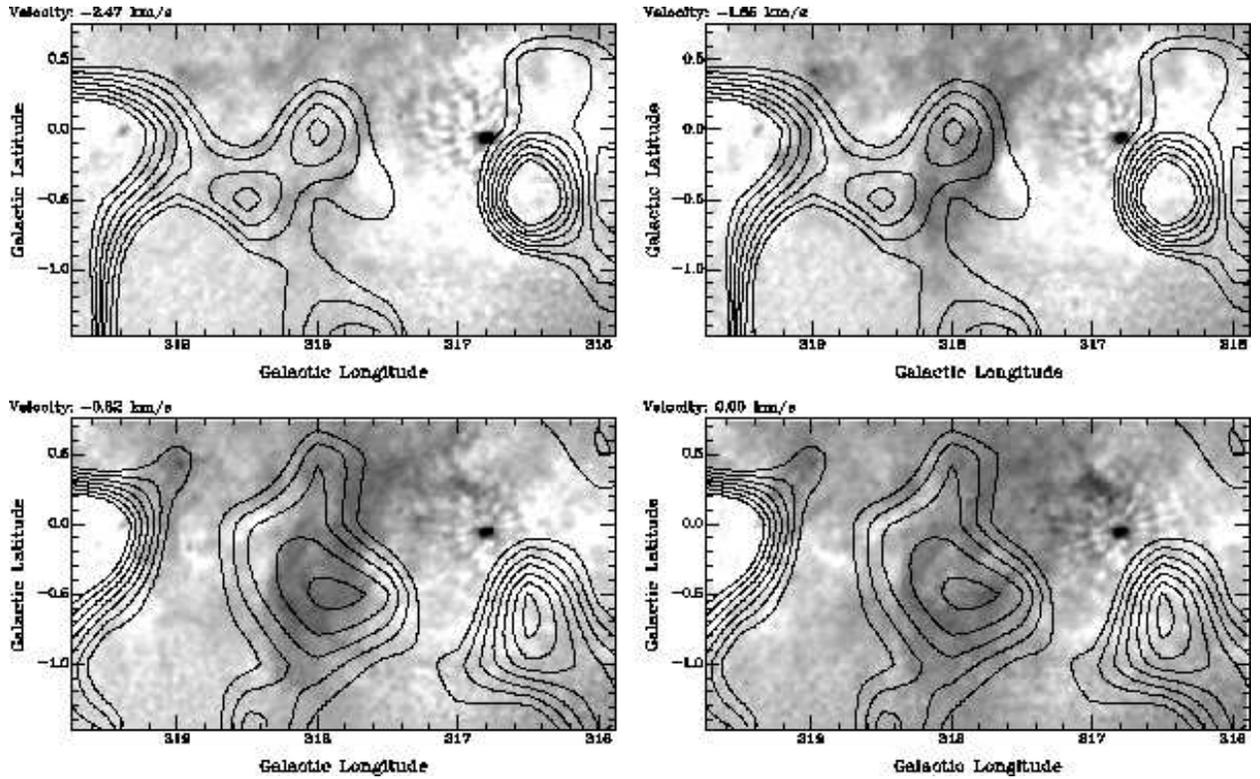}
\caption{Four images of the HISA cloud at velocities ranging from -2.5 $km$ $s^{-1}$ to 0.0 $km$ $s^{-1}$ with overlaid CO emission contours.  The contour levels range from 0.1 to 0.4 K in increments of 0.05 K.  The CO velocities are -1.95 $km$ $s^{-1}$ for the upper maps and -0.64 $km$ $s^{-1}$ for the lower maps.  The peak CO brightness temperature at the center of the HISA cloud is 0.3 K.}
\end{figure}

\end{document}